\documentclass[preprint,12pt]{elsarticle}

\usepackage{amsmath,amssymb}

\journal{latex template}

\begin{document}

\begin{frontmatter}



\title{Role of isospin and its conservation in neutron-rich fission fragments}


\author[rvt]{Swati Garg\corref{cor1}}
\cortext[cor1]{Corresponding author}
\ead{swat90.garg@gmail.com}
\author[rvt,focal]{Bhoomika Maheshwari}
\author[rvt]{Ashok Kumar Jain}
\address[rvt]{Department of Physics, Indian Institute of Technology Roorkee, Roorkee-247667, India}
\address[focal]{Department of Physics, Banasthali Vidyapith, Banasthali-304022, India.} 

\begin{abstract}
Following upon our earlier paper~\cite{swati} containing some initial results, we present detailed discussion and complete results in this paper, which provide the first direct evidence for the validity of isospin as a nearly good quantum number in neutron-rich systems. The evidence comes from the reproduction of the general features of the partition-wise relative yields of neutron-rich fission fragments produced in two heavy-ion induced fusion fission reactions, namely $^{208}$Pb ($^{18}$O, f) and $^{238}$U ($^{18}$O, f), by using the concept of isospin conservation. To fix the isospin values and use the isospin algebra, we invoke what we term as Kelson's conjectures. We present a consistent scheme for isospin assignments based on these considerations. Our calculated results confirm that isospin behaves as an approximately good quantum number in neutron-rich systems, in this case, the fission fragments.
\end{abstract}


\end{frontmatter}


\section{Introduction}
\label{intro}
Isospin is one of the most fundamental concepts in particle physics and is widely used to obtain the cross-sections in different isospin channels for a given reaction or, decay. The role of isospin in nuclear physics has always been thought of as rather less fundamental. For the first time, Wigner~\cite{wigner} introduced isospin as a fundamental concept in nuclear physics, which enables one to explore the complex nuclear structure in a simplified way. It may be noted that isospin is generally considered to be applicable to light nuclei where isospin mixing is small due to small Coulomb interaction~\cite{warner}. Nearly 50 years ago, Robson~\cite{robson} in his review on isospin suggested that isospin can play a very significant role in understanding many nuclear phenomena, especially analog resonances in heavy nuclei. In another detailed review of isospin in nuclear physics, Temmer~\cite{temmer} has also discussed the goodness of isospin in heavy nuclei and concluded that ``Isospin as a good quantum number is a much simpler hypothesis and useful until shown to be inadequate". Bohr and Mottelson~\cite{bohr} have also discussed this idea and concluded that even in heavy nuclei, the Coulomb potential mixes higher isospin states with a very small probability in the ground state. Robson again reiterated in 1973 ~\cite{robson1} that ``Isospin has been reborn as an important and useful quantum number for all nuclei" including heavy nuclei.

In 1960's, French and MacFarlane~\cite{french} predicted the usefulness of isospin in reactions where a single nucleon is added or removed from the target nucleus to study the isospin splitting of resonances. Isobaric Analog States (IAS), which are states in neighboring isobars whose isospin (and of course spin and parity) are same, played an important role in investigating the goodness of isospin in heavier nuclei. In one of the first such evidences, Anderson ${et}$ ${al.}$~\cite{anderson} experimentally showed that in a (p,n) reaction, the final product is found in a state which is IAS of the corresponding state in the target nucleus. It was observed that the width of the peak in the neutron energy spectrum corresponding to this state is quite small suggesting a $\Delta T=0$ population, and also that isospin is possibly a good quantum number. Fox ${et}$ ${al.}$~\cite{fox} further expanded the work and observed analog states as compound nucleus resonances. The resonances observed in proton induced reactions and (d, p) reactions show similar behavior. The first detailed theoretical work on isospin conservation in heavy nuclei came from Lane and Soper in 1962~\cite{lane}. Similar other papers~\cite{bohr, sliv} along with the reviews of Robson~\cite{robson}, Temmer~\cite{temmer}, and Auerbach~\cite{auerbach} put the concept of isospin back into focus for heavy mass nuclei.

A very brief account of our calculations and initial results for $^{208}$Pb ($^{18}$O, f) have recently been presented in references~\cite{swati, jain1}. In the present work, we present a detailed analysis of fission fragment mass distribution from three measurements~\cite{bogachev, banerjee, danu}. The measurements of Bogachev ${et}$ ${al.}$~\cite{bogachev}, Banerjee ${et}$ ${al.}$~\cite{banerjee} and Danu ${et}$ ${al.}$~\cite{danu} are the first of its kind where both $Z$ and $A$ are precisely identified, although only for even-even fission fragments. The fission fragment mass distributions were obtained for the compound nucleus $^{226}$Th and $^{256}$Fm formed in the reaction $^{18}$O+$^{208}$Pb and $^{18}$O+$^{238}$U respectively, employing the large detector arrays EUROBALL IV~\cite{simpson} and INGA~\cite{muralithar} by using fission fragment gamma ray spectroscopy. The relative intensities of gamma ray transitions between the levels of ground rotational bands of pair of light and heavy even-even fragments provide the relative yields of pair of fission fragments. All the even-even fragments in the range $A= 76-144$ and $A= 90-158$ for the two reactions $^{208}$Pb ($^{18}$O, f) and $^{238}$U ($^{18}$O, f) respectively, have been identified and intensities measured by gamma ray tagging of fragments. The pair of correlated fission fragments also provides information on neutron multiplicity which is also crucial to our calculations. 

In this paper, we present detailed calculations and results, which show that isospin is able to reproduce the main features of fission fragment mass distribution providing the first direct evidence of near goodness of isospin in neutron-rich systems~\cite{swati,jain1, jain}. However, no shell effects and corrections for isomeric transitions have been taken into account. We would like to stress that our work does not represent a model for calculating fission fragment distributions. We have rather used the fission fragment distribution data to support the goodness of isospin in neutron-rich nuclei. Further, these results may be valid only for compound nuclear fission data. 

We present in section~\ref{sec:1}, a discussion of the extent of isospin mixing in neutron-rich nuclei emitted in fission based on the earlier works~\cite{bohr, lane, sliv, auerbach}. Further, we present the conjectures of Kelson~\cite{kelson}, which are helpful in isospin assignments during the fission process. We present in section~\ref{sec:2} the experimental data and formalism for the calculation of relative yields of fission fragments emitted in $^{208}$Pb ($^{18}$O, f) and $^{238}$U ($^{18}$O, f) reactions. In section~\ref{sec:3}, we have discussed the results of our calculations and in section~\ref{sec:4}, we summarize the outcomes of this paper.

\section{Isospin Mixing and Assignments in neutron-rich systems}
\label{sec:1}
A nucleon is assigned the isospin quantum number $T=1/2$. Similar to spin, the three components of the isospin vector $T$ are denoted by $T_1$, $T_2$ and $T_3$. The value of $T_3$ distinguishes between a proton and a neutron: $T_3=+1/2$ for neutron and $-1/2$ for proton. The projection of total isospin of a ($N, Z$) nucleus is given by $T_3=(N-Z)/2$. The maximum isospin of a ($N, Z$) nucleus can be $T=(N+Z)/2$ and the minimum can be $(N-Z)/2$. It is also generally accepted that the ground state of a nucleus carries the lowest isospin i.e. $T=(N-Z)/2$. However, there is no simple prescription available so far to assign the total isospin $T$ to other nuclear states. 

It is known that the total isospin $T$ is conserved in strong interactions whereas only the third component of isospin $T_3$ is conserved in the electromagnetic interactions. Since $Z$ becomes very large in heavy nuclei, the impurities due to isospin mixing may increase and isospin conservation may be expected to be badly broken. So, for the light nuclei ($A<40$), isospin may remain a reasonably good quantum number and its conservation is considered to be valid and studied~\cite{warner}. However, it is possible that the isospin may remain quite pure in neutron rich heavy nuclei also because the excess neutrons dilute the isospin impurity of the core ($N=Z$) to give a quite pure isospin as shown in the theoretical work of Lane and Soper~\cite{lane}. 

\subsection{Purity of Isospin in neutron-rich nuclei}
Lane and Soper~\cite{lane} showed that isospin is a good quantum number in heavy nuclei based on the numerical estimates of isospin mixing due to the Coulomb potential. While the higher lying $T=T_0$ states can heavily mix into the $T_0$ ground state, Lane and Soper showed that the mixing of $T=T_0+1$ states decreases by a factor of $2(N-Z+2)^{-1}$. Therefore, as we move towards neutron-rich nuclei, the isospin mixing between states with different isospins decreases drastically.
 
The total isospin of a neutron-rich nucleus may, therefore, be looked upon as the sum of isospin of the ($N=Z$) core  to the isospin of the ($N-Z$) excess neutrons. The isospin of the core is mainly zero but may have some impurity by mixing of one unit higher isospin due to Coulomb interaction. The addition of excess neutrons having a pure isospin of $(N-Z)/2$ can, therefore, produce a total isospin of $(N-Z)/2$ or $(N-Z+2)/2$. The admixture of $(N-Z+2)/2$ isospin with $(N-Z)/2$ gets reduced by a factor of $2/(N-Z+2)$ resulting in a relatively pure isospin in heavy nuclei which are neutron-rich~\cite{lane}. Primary fission fragments produced in heavy ion fusion fission reactions have a large $N/Z$ ratio which is similar to that of the compound nucleus ($CN$). Even the secondary fission fragments have nearly similar $N/Z$ ratios and even if one member of the pair of fragments is neutron-rich, it may ensure purity of isopsin. Therefore, the large isospin of the $CN$ carries over into the fission fragments and may remain an approximately good quantum number.

Sliv and Kharitonov also calculated the isospin mixing of one unit higher isospin state on the basis of the independent particle model~\cite{sliv}. They estimated that along the $\beta$-stability line, isospin mixing first increases to 7\% upto $^{40}$Ca ($N=Z$) and then starts decreasing as neutron number exceeds the proton number, reducing to 2\% for $^{208}$Pb. Bohr and Mottelson~\cite{bohr} have used the hydrodynamical model to calculate the isospin mixing and obtained a very small mixing in $N>Z$ nuclei as compared to the $N=Z$ nuclei. Auerbach~\cite{auerbach} in his review has compared the isospin mixing from various approaches like the shell model, harmonic oscillator, hydrodynamical model, RPA etc. He concluded that the isospin impurity gets diluted as ($N-Z$) increases on the basis of the RPA results, which are considered to be most reliable. 

\subsection{Kelson's conjectures}
The assignment of total isospin to fission fragments is a rather difficult task as there is no standard prescription for doing so. We find the beautiful arguments presented by Kelson very useful in this regard~\cite{kelson}. Kelson suggested that the isospin may play a crucial role in fission phenomenon and thus fission can be used to explore the IAS~\cite{kelson}. From the liquid drop model (LDM) mass formula, the binding energy for the ground state of a ($N, Z$) nucleus with mass number $A$ is given in terms of the volume, surface, Coulomb, asymmetry and pairing energies. The mass formula needs to be generalized to obtain the binding energy of each possible shape that the system would acquire during a dynamic process like fission.

We know that fission occurs mainly because of the competition between the surface and Coulomb energy terms. But we should also take into account the role of asymmetry term in fission. Being proportional to isospin $T_3^2$ in ground state, it becomes proportional to $T^2$ as $T=T_3$ in the ground state. On generalizing it for the excited states populated during fission, the value of this term will change because of the increase in T as the nucleus goes to higher excited states.

Kelson then invoked the independent particle model (IPM) picture to bring in the microscopic details. If it is assumed that the neutron and proton wave functions are identical then the occupation of the lowest orbits implies that that the isospin remains conserved as the system evolves through various shapes and the asymmetry term may be neglected. However, the assumption of occupying lowest orbits is not justified in a dynamic process like fission. This may have important consequences on isospin as pointed out by Kelson in the following.

Due to occupation of higher excited states, the asymmetry energy term which is proportional to $T^2$, no longer remains a constant. We can also say that in the presence of strong Coulomb forces, the expectation value $\langle T^2 \rangle$  in the Hamiltonian gives the degree of isospin non-conservation. Kelson next considered the expectation value of $T^2-T_3(T_3+1)$ instead of $T^2$. If the system is in ground state, then this expectation value will be zero as $T=T_3$ for the ground state. This expectation value will give us a kind of deviation in energy when we move from ground state to higher excited states, as is expected during fission phenomenon. For an independent particle wave function $\phi$ consisting of $Z$ protons and $N$ neutrons, distributed in the same complete set of single particle basis for both protons and neutrons, it follows that
\begin{equation}
N-Z = \nu-\pi 
\end{equation}
where $\nu$ denotes the number of neutrons in orbits unoccupied by protons and $\pi$ denotes the number of protons in orbits unoccupied by neutrons. Using angular momentum algebra, one obtains
\begin{equation}
\langle \phi \mid T^2-T_3 (T_3+1) \mid \phi \rangle = \langle \phi \mid T_- T_+ \mid \phi \rangle = \pi 
\end{equation}
Since the electrostatic repulsion acts between protons only, Kelson suggested that there is a gradual increase of $\pi$ and hence of $T>T_3$ components during the process. 

After fission, the system is divided into two subsystems characterized by $\pi_i$ such that $\pi=\pi_1+\pi_2$. The fission fragments lose maximum part of their energy in neutron emission. Further, neutron emission will either increase $\pi$ value or keep it same. Therefore, fission fragments are more likely to be formed in excited states with $T>T_3$. Kelson, further proposed that ``the tendency to overpopulate highly excited states with $T>T_3$ in the primary fission products, carries largely over to the conventionally referred to Isobaric Analog states (IAS) in the observed products". 

These observations by Kelson may be summarized as: i) The neutron emission during fission process enhances the formation of higher excited states in fission fragments with $T>T_3$, and ii) The observed fission fragments would preferably be formed in IAS. We term both these observations as Kelson's conjectures and find them very useful in assigning the total isospin of the fission fragments.

\section{Experimental data and Formalism}
\label{sec:2}
\subsection{Experimental data}
We give a brief description of the experimental data used by us, which is so crucial to our calculations. We have considered two reactions in this paper, namely $^{208}$Pb($^{18}$O, f) and $^{238}$U($^{18}$O, f), where fission fragmnet distribution data have been measured with a precision of one unit of $Z$ and $A$, giving partition wise fission fragment yields for even-even nuclei only. Two sets of experimental data are known for $^{208}$Pb($^{18}$O, f), which come from the measurements of Bogachev ${et}$ ${al.}$~\cite{bogachev} and Banerjee ${et}$ ${al.}$~\cite{banerjee}. This reaction leads to the formation of $^{226}$Th as CN. Both of them have reported the experimental observation of six distinct partitions namely Ru-Pd, Cd-Te, Zr-Sn, Sr-Te, Kr-Xe and Se-Ba. While Banerjee ${et}$ ${al.}$~\cite{banerjee} have observed 59 fragments in total, Bogachev ${et}$ ${al.}$~\cite{bogachev} have observed 65 fragments. Bogachev ${et}$ ${al.}$~\cite{bogachev} have also reported the neutron multiplicity data for each of the six partitions and observed that the 6$n$ emission channel has maximum number of counts in all the partitions except in Se-Ba partition where the 4$n$ emission channel dominates. However, Banerjee ${et}$ ${al.}$~\cite{banerjee} have reported the neutron multiplicity data for first four partitions only i.e. Ru-Pd, Cd-Te, Zr-Sn and Sr-Te. The 6$n$ emission channel is again observed to be the most dominating channel for these partitions. The average neutron multiplicity data is nearly consistent from the two measurements, considering the experimental uncertainties quoted by Banerjee ${et}$ ${al.}$~\cite{banerjee}. However, the experimental data of Bogachev ${et}$ ${al.}$~\cite{bogachev} is more complete and the multiplicity data extend from zero to 12-14 neutron emission. Therefore, we have used the neutron multiplicity data from Bogachev ${et}$ ${al.}$~\cite{bogachev} in our calculations.

Danu ${et}$ ${al.}$~\cite{danu} have reported similar kind of experimental data for the reaction $^{238}$U($^{18}$O, f), again for even-even fragments only. The CN formed in this case is $^{256}$Fm, which is quite different from the earlier case. The authors have reported the fragment yields for 65 fragments in seven partitions, namely, Sn-Sn, Cd-Te, Pd-Xe, Ru-Ba, Mo-Ce, Zr-Nd and Sr-Sm. The neutron multiplicity data are not reported by the authors. However, the dominating $n$-emission channels for each of the seven partitions are given in the paper. For Sn-Sn partition, $12n$ emission channel is the dominating channel and for Ru-Ba partitions, $8n$ channel is the dominating one. For rest of the five partitions, $10n$ emission channel is the dominating $n$-emission channel. This information has been used by us in our calculations to define the isospin of the residual compound nucleus. 

\subsection{Assignment of total isospin to the fission fragments}
Assignment of the total isospin to a state in which fission fragments are formed is not so straightforward. The projection of isospin $T_3$ is, however, always known. Kelson's conjectures guide us in this regard. Let us consider a compound nucleus ($CN$) formed in a heavy ion fusion reaction which fissions into two fragments $F_1$ and $F_2$ along with the emission of $n$ number of neutrons. This process may be depicted as,
\begin{equation}
Y(T_Y,T_{3_{Y}})+X(T_X,T_{3_{X}}) \rightarrow CN(T_{CN},T_{3_{CN}})\rightarrow F_1(T_{F1},T_{3_{F1}})+F_2(T_{F2},T_{3_{F2}})+n 
\end{equation}
where $T_Y, T_X, T_{CN}, T_{F1}$ and $T_{F2}$ are the total isospin values of projectile, target, $CN$, and the two fragments respectively. Conservation of the third component $T_3$ implies that $T_{3_{Y}}+T_{3_{X}}=T_{3_{CN}}$. From the isospin algebra, we have 
\begin{equation}
\mid T_{X}-T_Y \mid \leq T_{CN} \leq (T_{X}+T_{Y})
\end{equation} 
We assume that the target and projectile are in their ground states and the total isospin $T$ for the ground state of a nucleus is equal to its minimum value, i.e., $T=T_3$. Thus, we have $T_Y=T_{3_{Y}}$ and $T_X =T_{3_{X}}$. Therefore, we obtain 
\begin{equation}
\mid T_{3_{X}}-T_{3_{Y}}\mid \leq T_{CN} \leq (T_{3_{X}} +T_{3_{Y}})
\end{equation}
But the total isospin is $T_{CN}\geq T_{3_{CN}}$ where $T_{3_{CN}}=T_{3_{X}}+T_{3_{Y}}$. Thus, the only possible value of $T_{CN}=T_{3_{X}}+T_{3_{Y}}$. This gives us a unique value of isospin for the $CN$. 
For example, in the heavy ion induced reaction $^{208}$Pb ($^{18}$O, f) being considered in this paper, the projectile and target are initially assumed to be in the ground state. The target and projectile, therefore, have
\begin{equation}
T_X(^{208}Pb)= T_{3_{X}}(^{208}Pb)=22, T_Y(^{18}O)=T_{3_{Y}}(^{18}O)=1
\end{equation}
From the conservation of isospin, the compound nucleus $^{226}$Th can have three possible values of $T_{CN}$= 21, 22, 23. However, $^{226}$Th, which is the CN, has $T_{3_{CN}} = 23$. Since $T_{CN} \geq T_{3_{CN}}$, $T_{CN}$ can have only one value, i.e., 23. This uniquely fixes the total isospin of the $CN$.

The $CN$ now fissions into a pair of daughter fragments with the emission of $n$ number of neutrons. For simplicity, we assume that all the neutrons that are emitted during the fission process are released in one step and do not make any distinction between the pre and post scission neutrons because the time difference between the two is of the order of 10$^{-19}$ sec which is very small. It should not make any impact on the final results as such as we are not concerned with the dynamics of fission in our work. It also allows us to define a residual compound nucleus (RCN) that remains after the emission of $n$ neutrons from $CN$. This simplifies the procedure of isospin assignments and reduces the many body problem in fission to a simple two body problem. The total isospin of the $RCN$ will, therefore, lie in the range,
\begin{equation}
\mid T_{CN}-n/2 \mid \leq T_{RCN} \leq (T_{CN}+n/2)
\end{equation}
and $T_{3_{RCN}}=T_{3_{F1}}+T_{3_{F2}}=T_{CN}-n/2$. Alternatively, $T_{RCN}$ should also satisfy,
\begin{equation}
\mid T_{F1}-T_{F2} \mid \leq T_{RCN} \leq (T_{F1}+T_{F2})
\end{equation}
As we shall see later, both the equations must be satisfied simultaneously, and this limits the range of $T_{RCN}$ values allowed in the calculation.

Using Kelson's first conjecture, neutron emission will lead to the formation of states with $T>T_3$ in the fission fragments and, therefore, the RCN also which is an auxiliary system composed of the two fragments. We, therefore, choose the maximum of the allowed values of $T_{RCN}$ i.e. $T_{RCN}=T_{F1}+T_{F2}$ with the condition that Eq. (7) is also satisfied.

Another crucial task is to assign total isospin values to the fission fragments. We now invoke Kelson's second conjecture according to which the fission fragments are preferably formed in IAS. The isobars, where the IAS are formed, constitute a multiplet. 

We limit ourselves to three isobars corresponding to each mass number having $T_3$ values as $T'_3$, $T'_3$+2, and $T'_3$+4. For each such isobaric triplet corresponding to a particular mass number, we assign the total isospin value $T=T'_3+4$ which is the maximum of the three $T_3$ values. These fragments are formed in IAS which decay to the ground state by the emission of $\gamma$-rays only, leading to no change in $T_3$ value although $T$ value can change during $\gamma$-emission as allowed by the selection rules. We note that the maximum contribution to the yield anyway comes from the fragments having the maximum isospin projection for a given isospin value. Of the three states in our hand, i.e. ($T=T'_3+4$, $T_3=T'_3+4$), ($T=T'_3+4$, $T_3=T'_3+2$), and($T=T'_3+4$, $T_3=T'_3$), the first one will lie lowest in energy while the rest two will have higher excitation energy. The contribution of those states, which have smaller $T_3$ values, decreases naturally. Thus, the fission fragments are primarily formed in the maximum projection states. If we consider a large number of isobars, the multiplet will be larger and lead to a larger $T$ value; such large $T$ states are practically not feasible to populate in most of the members of the multiplet. We find that taking three members in an isobaric multiplet is sufficient.  

\subsubsection{Assignments of isospin for $^{208}$Pb ($^{18}$O, f)}

Let us now consider the assignment of isospin values to the fission fragments in all the six partitions of the reaction $^{208}$Pb ($^{18}$O, f). The CN formed in this reaction is $^{226}$Th. The measurements were made for six partitions, namely Ru-Pd, Mo-Cd, Zr-Sn, Sr-Te, Kr-Xe and Se-Ba. The observed fragments and the neutron multiplicity data allows us to fix the isospin of the RCN and the fragments.  As discussed above, we choose three isobars for each mass number of the fission fragments. Further, we consider eight lighter and eight heavier fragments in each partition. As an example, we choose three isobars of $A=112$ to be $^{112}$Ru ($T_3=12$), $^{112}$Pd ($T_3=10$) and $^{112}$Cd ($T_3=8$). We, therefore, assign $T=12$ for $A=112$ fragments, which can give us all the three projections namely, $T_3$=12, 10, and 8. The contribution of ($T=12$, $T_3=12$) to the yield will be highest, while ($T=12$, $T_3=10$) and ($T=12$, $T_3=8$) which lie higher in excitation energy, contribute very little to the yield for $A=112$. 

\begin{figure}
\includegraphics[width=15cm,height=12cm]{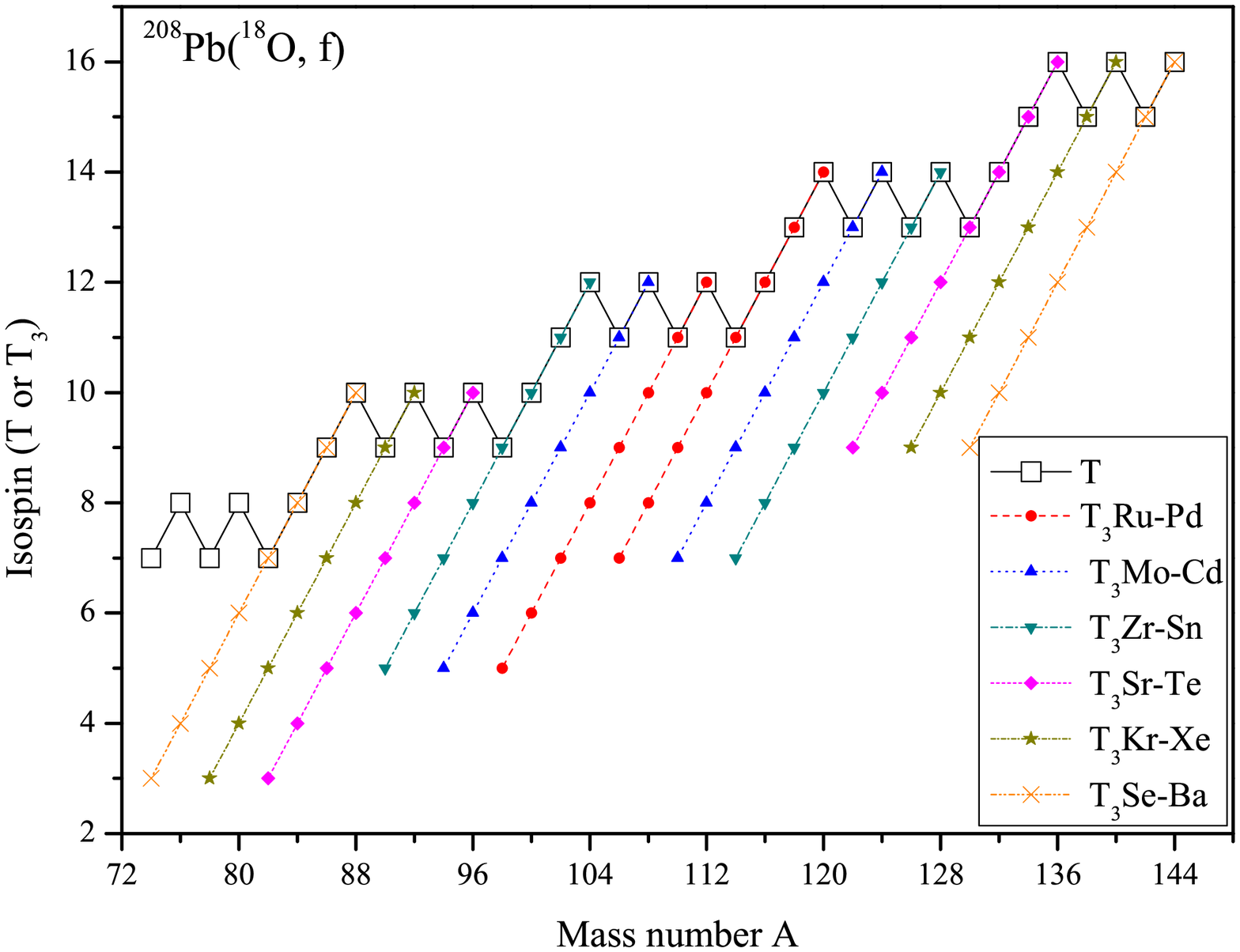}
\caption{\label{fig:tt3}(Color online) Isospin $T$ or $T_3$ vs. mass number $A$ of the final fission fragments emitted in $^{208}$Pb ($^{18}$O, f) after the emission of neutrons. Open squares on the zig-zag line show the isospin $T$ assigned to each mass number. Other symbols show the $T_3$ values for the fragments of different partitions. One particular type of symbol corresponds to $T_3$ values for the fragments of a distinct partition. Out of the two lines connecting the same symbols, the one on the right hand side is for the heavier and the one on the left hand side is for the lighter fragments.}
\end{figure}

Following this procedure, we assign total isospin to all the isotopes. As an example, we show the assignments made for Ru isotopes in Table~\ref{tab:t_ru}. There are three isobars for each mass number from $A=98$ to $A=112$. The assigned $T$ values are shown in last column of Table~\ref{tab:t_ru}. In a similar way, we can assign the isospin to all the fragments emitted in $^{208}$Pb($^{18}$O, f) as depicted in Fig.~\ref{fig:tt3}. This prescription allows us to assign the isospin values in a systematic and consistent manner. We can see from the figure that there are three sets of eight nuclides in the middle and two sets of six nuclides at the two extremes (open square symbols). Also, the six partitions are shown by the six pairs of inclined straight lines having same symbols as shown in the inset. Here, we are considering only the experimentally observed partitions but if we go beyond these partitions, the last two sets may also have eight nuclides. 

\begin{table}
 \caption{The table lists the $T$ values assigned to all the isotopes of Ru from $A=98$ to $A=112$, by considering three isobars for each mass number.}
 \label{tab:t_ru}
 \begin{center}
 \begin{tabular}{|c|c|c|c|c|c|c|c|}
 \hline
 A & Nucleus & $T_3$ & Nucleus & $T_3$ & Nucleus & $T_3$ & $T$ \\ 
 \hline
  \rule{0pt}{10pt} 98 & $^{98}$Ru & 5 & $^{98}$Mo & 7 & $^{98}$Zr & 9 & 9 \\
 \hline
  \rule{0pt}{10pt}100 & $^{100}$Ru & 6 & $^{100}$Mo & 8 & $^{100}$Zr & 10 & 10 \\
 \hline
  \rule{0pt}{10pt}102 & $^{102}$Ru & 7 & $^{102}$Mo & 9 & $^{102}$Zr & 11 & 11 \\
 \hline
  \rule{0pt}{10pt}104 & $^{104}$Ru & 8 & $^{104}$Mo & 10 & $^{104}$Zr & 12 & 12 \\
 \hline
  \rule{0pt}{10pt}106 & $^{106}$Pd & 7 & $^{106}$Ru & 9 & $^{106}$Mo & 11 & 11 \\
 \hline
  \rule{0pt}{10pt}108 & $^{108}$Pd & 8 & $^{108}$Ru & 10 & $^{108}$Mo & 12 & 12 \\
 \hline
  \rule{0pt}{10pt}110 & $^{110}$Cd & 7 & $^{110}$Pd & 9 & $^{110}$Ru & 11 & 11 \\
 \hline
  \rule{0pt}{10pt}112 & $^{112}$Cd & 8 & $^{112}$Pd & 10 & $^{112}$Ru & 12 & 12 \\
 \hline
 \end{tabular}
\end{center}
\end{table}

We, then, calculate the isospin of $RCN$ after the emission of $n$ number of neutrons just by adding the isospin values of the two fragments as $T_{RCN}=T_{F1}+T_{F2}$ and $T_{3_{RCN}}=T_{3_{F1}}+T_{3_{F2}}$ following Kelson's first conjecture. From Table~\ref{tab:n_trcn}, we note that the value of $T_{RCN}$ varies from 20 to 25 for different $n$-emission channels. From the experimental neutron multiplicity data~\cite{bogachev}, we know that 4$n$ and 6$n$ emission channels dominate over the other channels. Therefore, major contribution should come from $T_{RCN}=22, 23, 24, 25$.

\begin{table}
 \caption{The table lists the isospin of $RCN$, $T_{RCN}$ calculated for different $n$-emission channels in all the six partitions.}
\label{tab:n_trcn}
\begin{center}
 \begin{tabular}{|c|c|c|c|c|c|}
 \hline
 \rule{0pt}{8pt}$n$ & 0, 8 & 2, 6, 10 & 4 & 12 & 14\\ 
 \hline
 \rule{0pt}{10pt}$T_{RCN}$ & 23 & 22, 24 & 23, 25 & 21, 23 & 20, 22, 24 \\
 \hline
 \end{tabular}
\end{center}
\end{table}

For example, we consider the $12n$ emission channel in Ru-Pd partition. Here, $T_{3_{RCN}}=23-6=17$. The isospin of $RCN$ is fixed by adding the isospin values of the two fragments, $T_{RCN}=T_{F1}+T_{F2}$ (see Table~\ref{tab:tf1_tf2}) but it should also lie in the range 23$-$6,....,23+6 i.e. 17,....,29 from Eq. (7) and if it does not, then we exclude that particular combination. We obtain 6 possible combinations for Ru-Pd partition as listed in  Table~\ref{tab:tf1_tf2}. We carry out similar exercise for all the fragments in each partition.

\begin{table}
 \caption{The list of the $T$ and $T_3$ values assigned to all the possible combinations of fragments emitted in $12n$ emission channel of Ru-Pd partition.}
 \label{tab:tf1_tf2}
 \begin{center}
 \begin{tabular}{|c|c|c|c|c|c|c|}
 \hline
 Nucleus & $T_{F1}$ & $T_{3_{F1}}$ & Nucleus & $T_{F2}$ & $T_{3_{F2}}$ & $T_{RCN}$= \\&&&&&&$T_{F1}$+$T_{F2}$ \\ 
 \hline
 \rule{0pt}{10pt} $^{98}$Ru & 9 & 5 & $^{116}$Pd & 12 & 12 & 21 \\
 \hline
 \rule{0pt}{10pt}$^{100}$Ru & 10 & 6 & $^{114}$Pd & 11 & 11 & 21 \\
  \hline
 \rule{0pt}{10pt}$^{102}$Ru & 11 & 7 & $^{112}$Pd & 12 & 10 & 23 \\
  \hline
 \rule{0pt}{10pt}$^{104}$Ru & 12 & 8 & $^{110}$Pd & 11 & 9 & 23 \\
  \hline
 \rule{0pt}{10pt}$^{106}$Ru & 11 & 9 & $^{108}$Pd & 12 & 8 & 23 \\
  \hline
 \rule{0pt}{10pt}$^{108}$Ru & 12 & 10 & $^{106}$Pd & 11 & 7 & 23 \\
 \hline
 \end{tabular}
\end{center}
\end{table}

\subsubsection{Assignments of isospin for $^{238}U$ ($^{18}O$, f)}

\begin{figure}
\begin{center}
\includegraphics[width=15cm,height=12cm]{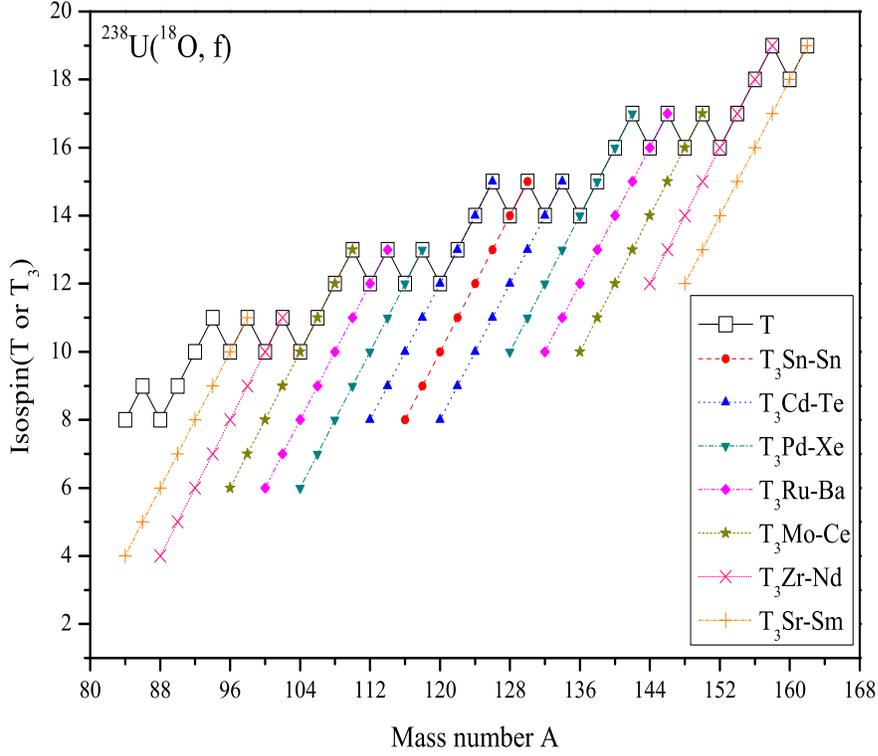}
\caption{\label{fig:tt3danu}(Color online) Similar to Fig.~\ref{fig:tt3} but for $^{238}U$ ($^{18}O$, f).}
\end{center}
\end{figure}

Let us now consider the second reaction $^{238}U$ ($^{18}O$, f), where the CN formed is $^{256}$Fm~\cite{danu}. The isospin of CN in this case is again unique and is given by $T_{CN}=T_{3_{CN}}=28$. The experimental measurements are given for 65 fragments in seven partitions in this reaction, namely, Sn-Sn, Cd-Te, Pd-Xe,Ru-Ba, Mo-Se, Zr-Nd, and Sr-Sm. The neutron multiplicity data are not given by the authors and we, therefore, rely on the information given for the dominant neutron emission channel to make our assignments. The dominant $n$-emission channel for Sn-Sn partition is 12$n$, Ru-Ba is 8$n$ and for the rest of the partitions, 10$n$ emission channel dominates. We assign the isospin value to various fission fragments emitted in this reaction using the prescription as discussed above for $^{208}$Pb ($^{18}$O, f). We again consider three isobars for each mass number of fission fragments and eight fragments each on the lighter side and the heavier side.  We also assign isospin to $RCN$ using Kelson's first conjecture as well as the Eq. (7) and Eq. (8), as explained above. The assigned isospin values to all the considered fission fragments are shown in Fig.~\ref{fig:tt3danu}.

\subsection{Calculation of relative intensities of fission fragments}
\label{sec:5}
After the assignment of isospin values, we proceed to calculate the relative intensities of neutron-rich fission fragments. First, we consider a particular $n$-emission channel of a given partition and construct all the possible pairs of fission fragments, satisfying the condition imposed by Eq. (8). We have restricted ourselves to the isospin part of the total wave function only. For each possible pair of fragments ($F_1, F_2$) identified by their isospins ($T_{F1} T_{3_{F1}}, T_{F2} T_{3_{F2}}$), emitted in $n^{th}$ neutron-emission channel in a particular partition, the isospin wave function of $RCN$ may be written as,
\begin{equation}
\mid{T_{RCN},T_{3RCN}}\rangle_n = \langle{T_{F1}T_{F2}T_{3_{F1}}T_{3_{F2}} \mid T_{RCN}T_{3_{RCN}}}\rangle\mid{T_{F1},T_{3_{F1}}}\rangle \mid{T_{F2},T_{3_{F2}}}\rangle
\end{equation}
where $\langle T_{F1}T_{F2}T_{3_{F1}}T_{3_{F2}} \mid T_{RCN}T_{3_{RCN}} \rangle$  represents the Clebsch-Gordon coefficient ($CGC$). The intensity of each pair of fragments in the respective partition for a particular $n$-emission channel is, therefore, given by,
\begin{equation}
I_n = \langle{CGC}\rangle^2 = \langle{T_{F1}T_{F2}T_{3_{F1}}T_{3_{F2}} \mid T_{RCN}T_{3_{RCN}}}\rangle^2
\end{equation}

We further introduce the weight factors for different $n$-emission channels as obtained from the neutron multiplicity distribution as given in Fig. 5 of Bogachev ${et}$ ${al.}$~\cite{bogachev}. It may be noted that the counts given in this figure are absolute counts. These values were read from the graph and normalized with respect to the $n$-emission channel having maximum number of counts in a given partition. Therefore, we obtain partition wise relative weight factors. The final yield of an individual fragment may now be obtained as,
\begin{equation}
I = \sum_{n} I_n \times w_n = \sum_{n} \langle{CGC}\rangle^2 \times w_n
\end{equation}
where the summation runs over all the neutron-emission channels reported by Bogachev ${et}$ ${al.}$~\cite{bogachev} and $w_n$ is the relative weight factor of a specific neutron-emission channel.

By using the isospin assignments as for example given in Table~\ref{tab:tf1_tf2}, we calculate their $CGC$ using Eq. (9) and square them. Then we multiply these by the values of the weight factors extracted from the experimental data to obtain the intensities. For example, the weight factor for the $12n$ emission channel in Ru-Pd partition is, $w_{12}$ = 0.0291. This is relative to the peak observed at 6$n$ emission channel which is normalized to unity.

Similarly, we carry out the calculation for all the $n$-emission channels in a given partition. Using Eq. (11), we then calculate the intensity of all the lighter as well as heavier mass fragments. We further normalize the yields of all the fragments with respect to the fragment having maximum yield, for the lighter and heavier set of fragments separately. This gives us the relative yields of fragments in a given partition which may be compared with the experimental data normalized in the same manner~\cite{bogachev, banerjee}. We repeat the same procedure as discussed above for all the partitions separately. These results are presented in the next section.

To calculate the total fission fragment mass distribution as a function of $A$, we simply add the individual yields of three isobars of each isobaric multiplet. The only difference here is in obtaining the weight factors, which obviously require experimental data. For this, we obtain the weight factors $v_n$ by normalizing with respect to the $n$-emission channel having maximum number of counts among all the partitions. These weight factors, therefore, relate to all the partitions. Thus, the total yield of a given mass fragment in a given partition may now be written as,
\begin{equation}
I' = \sum_{n} I_n \times v_n = \sum_{n} \langle{CGC}\rangle^2 \times v_n
\end{equation}
where $v_n$ is the relative weight factor for a particular $n$-emission channel. We calculate the yield of each and every fragment in all the six partitions. To calculate the total yield for a particular mass number, we add the yields of all the three isobars of the isobaric multiplet corresponding to that mass number. Finally, we calculate the relative yield for each mass number with respect to mass number having the maximum yield normalized to unity to obtain the total fission fragment mass distribution.

We now discuss the $^{238}$U ($^{18}$O, f) reaction. We know only the dominating $n$-emission channels in this case and the neutron multiplicity distribution data to obtain the relative weight factors of different $n$-emission channels have not been given by the authors. Therefore, in the calculations, we consider three $n$-emission channels centered around the dominant channel without the inclusion of any weight factors. For example, for Sn-Sn partition, $12n$ emission channel dominates, so we consider $10n, 12n$ and $14n$ channels in the calculations. Thus, modified equation for yields of fragments in a particular partition is given by,
\begin{equation}
I'' = \sum_{n} I''_n = \sum_{n} \langle{CGC}\rangle^2
\end{equation}
where the summation runs over all the $n$-emission channels under consideration. We repeat the same procedure as explained above to calculate the relative yields of fragments in all the seven partitions. In this case, we can not calculate the total fission fragment mass distribution as there are no relative weight factors for various partitions available. As a result, we can not say which partition will dominate over the others. This relative dominance will eventually decide the degree of symmetry/asymmetry in the total fission fragment mass distribution, which we can not predict.

\section{Results and Discussion}
\label{sec:3}
\subsection{$^{208}$Pb($^{18}$O, f)}
\begin{figure}
\begin{center} 
\includegraphics[width=13cm,height=14cm]{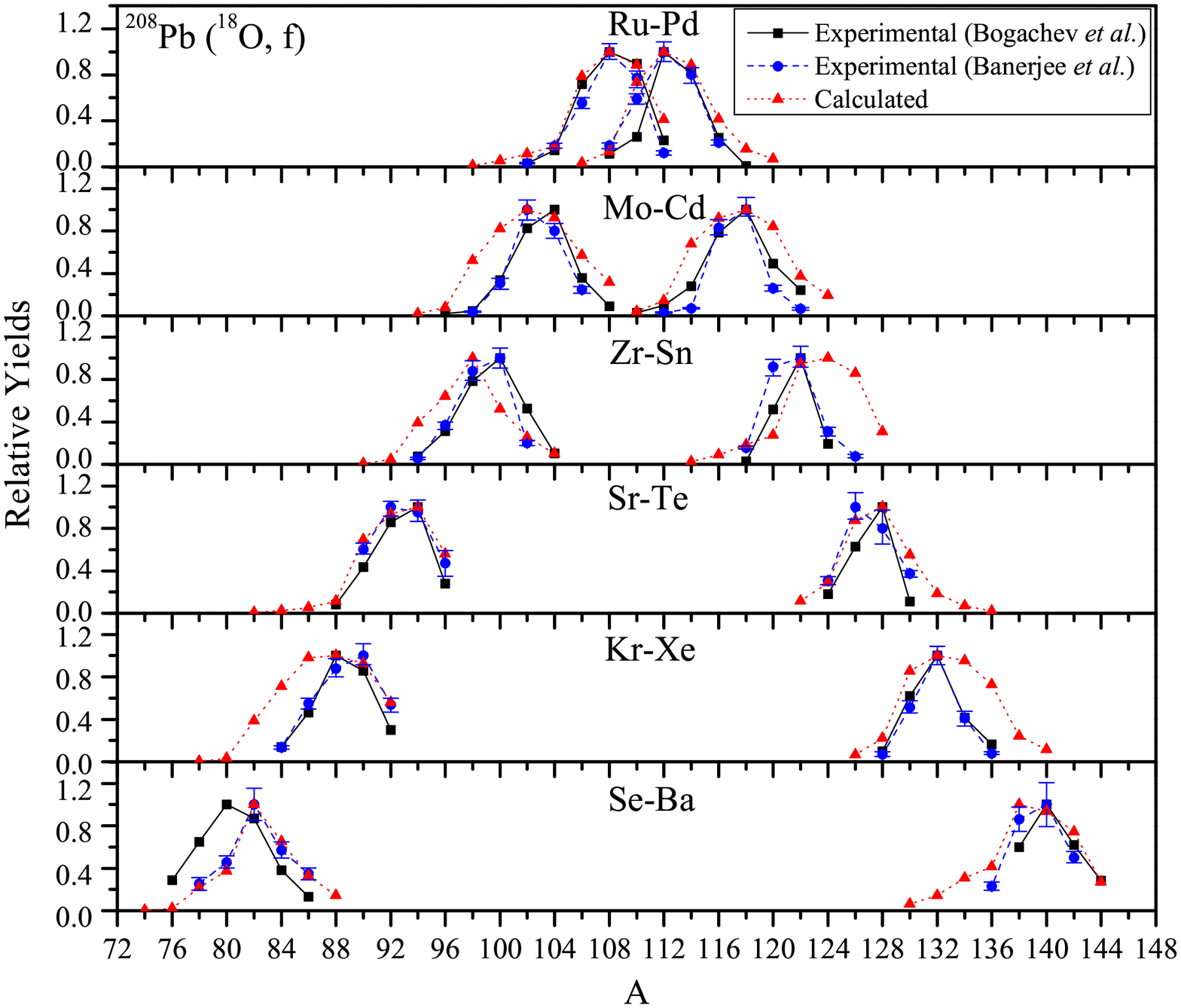}
\caption{\label{fig:bogabane1}(Color online) Comparison of the calculated and experimental relative yields of fission fragments vs. mass number $A$ of the final fission fragments formed after the emission of neutrons in all the six partitions of $^{208}$Pb ($^{18}$O, f). Experimental data are taken from Bogachev ${et}$ ${al.}$~\cite{bogachev} and Banerjee ${et}$ ${al.}$~\cite{banerjee}. In the calculated results, the weight factors from Bogachev ${et}$ ${al.}$~\cite{bogachev} have been used.}
\end{center}
\end{figure}

We, now, present the calculated results and their comparison with the experimental data. There are two sets of experimental data available for the reaction $^{208}$Pb($^{18}$O, f), one due to Bogachev ${et}$ ${al.}$~\cite{bogachev} and another due to Banerjee ${et}$ ${al.}$~\cite{banerjee}. However, we have used the neutron multiplicity data which has been reported by Bogachev ${et}$ ${al.}$~\cite{bogachev} only, as it is more complete. We also use the same normalization procedure for the experimental data to obtain relative yields in order to compare with the calculated results. We compare our results with both the measurements in Fig.~\ref{fig:bogabane1}. A reasonably good agreement between the experiment and theory is evident from Fig.~\ref{fig:bogabane1}. Since the error in the total yields has been estimated to be in the range of 10-30\% by Bogachev ${et}$ ${al.}$~\cite{bogachev}, the errors in the data for individual partitions will be at least 10\%.

We note that there are some differences between the two experimental data sets of Bogachev ${et}$ ${al.}$~\cite{bogachev} and Banerjee ${et}$ ${al.}$~\cite{banerjee}. For example, the peak in the relative yields occurs at $^{102}$Mo in Banerjee ${et}$ ${al.}$~\cite{banerjee} and at $^{104}$Mo in Bogachev ${et}$ ${al.}$~\cite{bogachev}. From Fig.~\ref{fig:bogabane1}, we see that the relative yields of at least three of the six partitions agree with the experimental data quite well. The relative yields of fission fragments in the partitions Ru-Pd, Mo-Cd and Sr-Te match very well with the experimental data of Bogachev ${et}$ ${al.}$~\cite{bogachev} On the other hand, the partitions Ru-Pd, Sr-Te and Se-Ba match reasonably well with the experimental data of Banerjee ${et}$ ${al.}$~\cite{banerjee}. Overall, we can say that among all the six partitions, there are only two partitions i.e., Zr-Sn and Kr-Xe whose experimental behavior is marginally different from what we calculate from theory. One possible reason for this deviation may be the shell effects due to the presence of $A=124$ ($Z=50$) and $A=136$ ($N=82$) closed shell. However, it is not clear how the closed shell will result in a dip in the observed mass distribution. It is, however, most likely due to a dip in the post neutron emission yields of the closed shell fragments which have higher excitation energies and large neutron separation energies. The shift in peak in calculations for Zr-Sn partition can also be attributed to the presence of an isomer in $^{98}$Zr (half-life, $t_{1/2}=1.7 \mu s$).

\begin{figure}
\begin{center} 
\includegraphics[width=13cm,height=14cm]{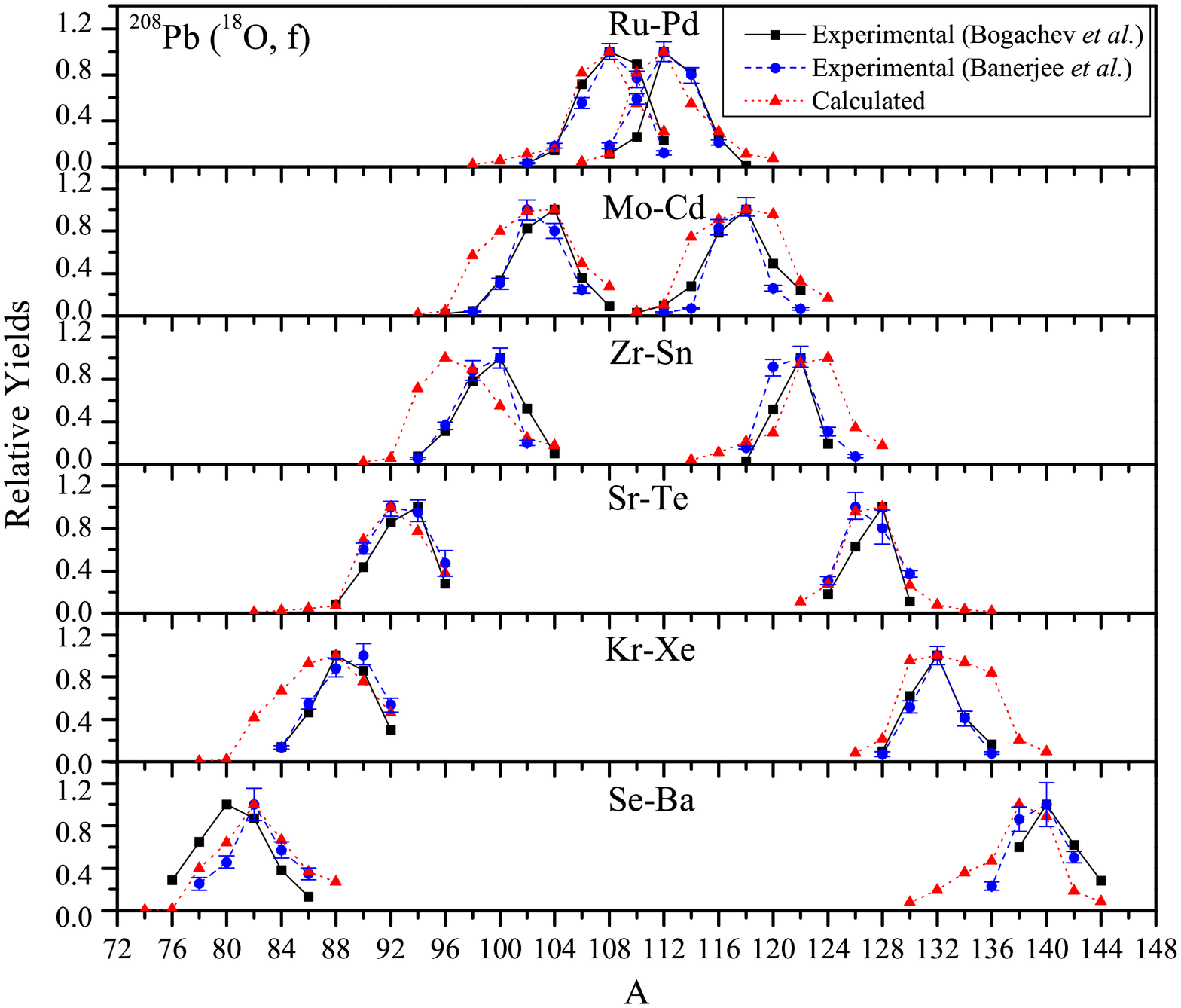}
\caption{\label{fig:bb468}(Color online) Similar to Fig.~\ref{fig:bogabane1} except that here we consider only $4n, 6n$ and $8n$ emission channels with the weight factors considered to be equal to unity.}
\end{center}
\end{figure}

The neutron multiplicity data reported by Bogachev ${et}$ ${al.}$~\cite{bogachev} may have an error of about 30\%. This may sometimes affect the calculated results adversely. We have verified that the results deteriorate slightly when all the weight factors are taken to be unity. We find that the most dominant channels corresponding to 4$n$, 6$n$, 8$n$ emissions are generally sufficient to reproduce the major features of the observed fragment distribution as shown in Fig.~\ref{fig:bb468}; these results differ marginally from those in Fig.~\ref{fig:bogabane1}. These results suggest that a calculation carried without any weight factors, when such data are not there, may also lead to reasonable estimates. This is the case of the results for $^{238}$U ($^{18}$O, f) reaction reported in the next subsection. We have also carried out a calculation by taking all the $CGC$'s equal to one to verify their role and check if weight factors of different $n$-emission channels alone are sufficient to explain the observed data. For $CGC=1$, the isospin does not play any role and we find that the calculated results deviate drastically becoming to broad to match the experimental data. This confirms that the fragment mass distributions are mainly controlled by the C.G. coefficients arising from isospin algebra.

Our isospin assignment may have an error of at the most one unit. We have verified that if we increase the isospin value of all the fragments by one unit, then the calculated results change negligibly. But, the results deteriorate completely  if T is changed by two units. Our isospin assignments are, therefore, reasonably good within an error of one unit.

\begin{figure}
\includegraphics[width=15cm,height=13cm]{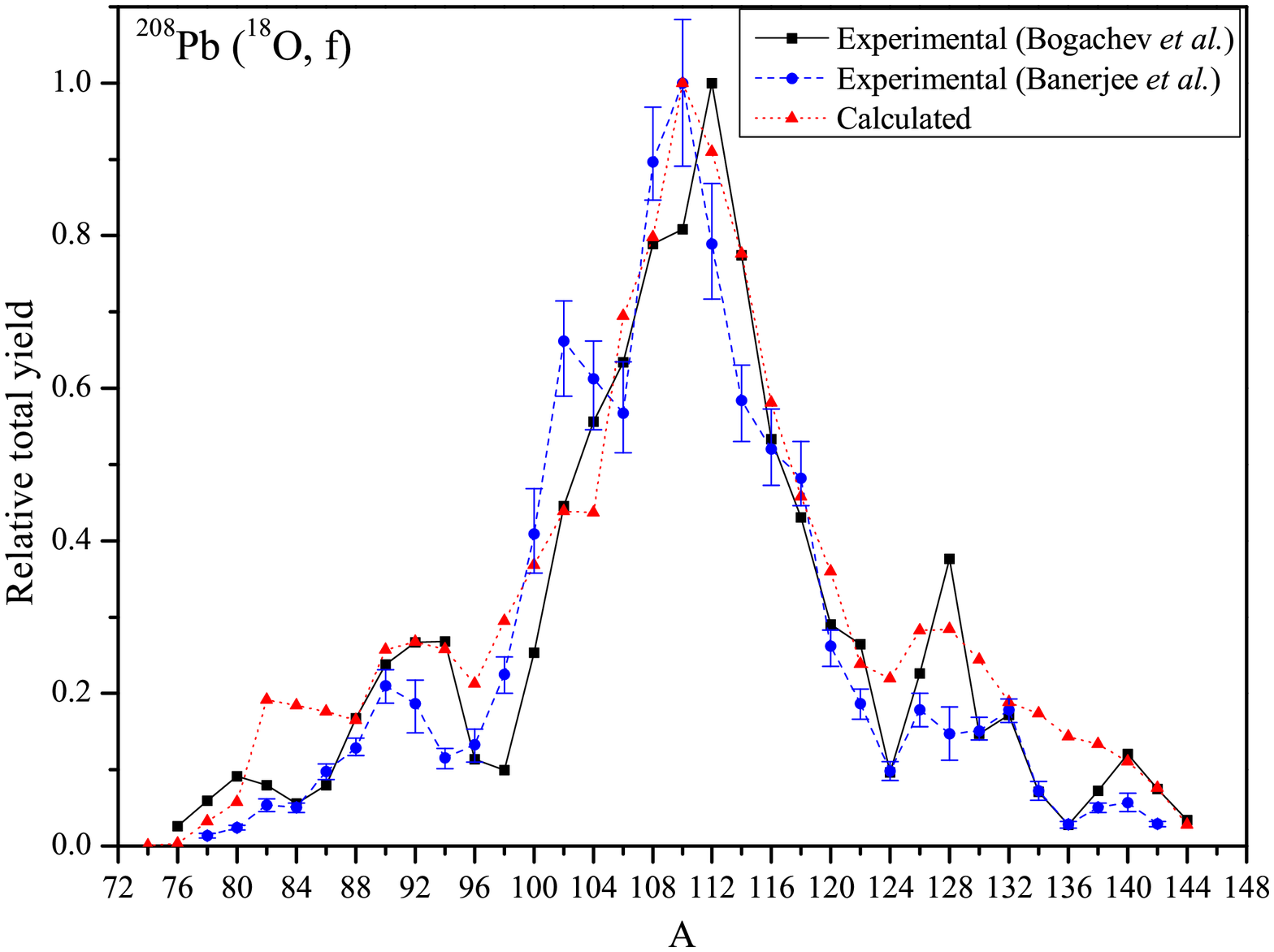}
\caption{\label{fig:bogabane2}(Color online) Comparison of the calculated and experimental total relative yields of fission fragments vs. mass number $A$ of the final fission fragments formed after the emission of neutrons for the reaction $^{208}$Pb ($^{18}$O, f). Experimental data are taken from Bogachev ${et}$ ${al.}$~\cite{bogachev} and Banerjee ${et}$ ${al.}$~\cite{banerjee}.}
\end{figure}

We have calculated the relative intensity for the total fragment mass distribution for $^{208}$Pb ($^{18}$O, f) as discussed in the section~\ref{sec:5} and plotted it with the experimental data as shown in Fig.~\ref{fig:bogabane2}. There is a reasonable agreement between the experimental and calculated values. The disagreement in the values at $A=124$ and 136 and the complementary fragments at $A=84$ and 98 in Fig.~\ref{fig:bogabane2} may be due to closed shell configuration as already discussed. It may also be pointed out that we have not considered any corrections due to the shell effects, the side-feedings and the presence of isomeric states. Even then, good agreements have been obtained with the data which confirms that isospin is reasonably pure and is a useful concept in neutron-rich systems.

\subsection{$^{238}$U ($^{18}$O, f)}
\begin{figure}
\begin{center} 
\includegraphics[width=13cm,height=14cm]{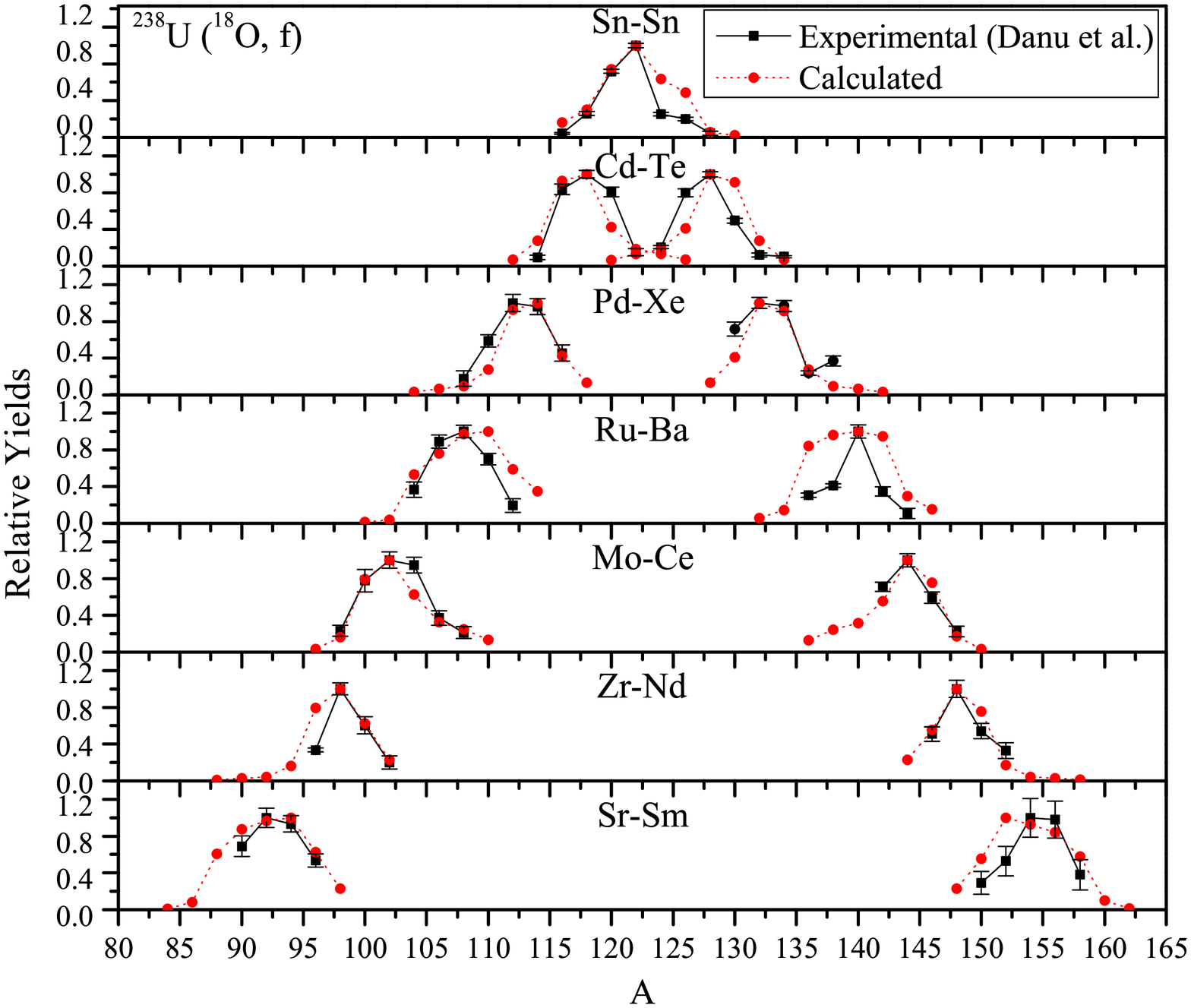}
\caption{\label{fig:danu}(Color online) Comparison of the calculated and experimental relative yields of fission fragments vs. mass number $A$ of the final fission fragments formed after the emission of neutrons in all the seven partitions of $^{238}$U ($^{18}$O, f). Experimental data are taken from Danu ${et}$ ${al.}$~\cite{danu}.}
\end{center}
\end{figure}

We, now, compare the partition-wise relative yields of fission fragments emitted in $^{238}$U ($^{18}$O, f) from our calculations with the experimental data from Danu ${et}$ ${al.}$~\cite{danu} in Fig.~\ref{fig:danu}. For Sn-Sn and Ru-Ba partitions, $12n$ and $8n$ channels are the dominating $n$-emission channels respectively and for all the other partitions, $10n$ emission channel is the dominating one. In the calculations, we consider only three channels centered around the dominating channel. For example, for the 12$n$ emission channel as the dominating channel, we consider 10$n$, 12$n$ and 14$n$ emission channels in the calculations. As already pointed out, even without any weight factors, it should be possible to reproduce the observed trends. In Fig.~\ref{fig:danu}, we can see that the calculated results for all the partitions except Ru-Ba partition follow the experimental data quite well. As discussed for $^{208}$Pb ($^{18}$O, f), there are shell effects at $A=124$ ($Z=50$) and $A=136$ ($N=82$) closed shell and their complementary fragments. Also, there are some long-lived isomers present in this mass region, e.g., $^{124}$Sn, $^{136}$Xe and $^{136}$Ba. These factors probably lead to the deviations between experiment and our results as we have not considered any of these effects in our calculations. 

\section{Conclusion}
\label{sec:4}
We have calculated the relative yields of neutron-rich fission fragments for all the six partitions of $^{208}$Pb ($^{18}$O, f) and seven partitions of $^{238}$U ($^{18}$O, f). We have also calculated the total fission fragment mass distribution of $^{208}$Pb ($^{18}$O, f). The calculated results have been compared with the three sets of experimental data available. These calculations have been carried out with a very simple approach based on isospin conservation without the inclusion of shell effects. These calculations were mainly driven by the expectations that the isospin behaves as approximately good quantum number for neutron-rich nuclei. This expectation is supported by the earlier theoretical works~\cite{bohr, lane, sliv, auerbach} which demonstrate that isospin becomes more pure as $(N-Z)$ value increases.  We found Kelson's conjectures~\cite{kelson} very helpful in assigning the total isospin to the fission fragments. Further, we have also used the neutron multiplicity data from Bogachev ${et}$ ${al.}$~\cite{bogachev} to include the weight factors of various $n$-emission channels in calculating the relative yields of fission fragments. The calculated values reproduce the experimental trends reasonably well for the mass distribution of fragments for both the reaction data sets, $^{208}$Pb ($^{18}$O, f) and $^{238}$U ($^{18}$O, f) in individual partitions and the total mass distribution of fragments for $^{208}$Pb ($^{18}$O, f). These results confirm that isospin remains reasonably pure in neutron-rich nuclei. 

We may emphasize again that we are in no way claiming to predict the fission fragment mass distribution. Our model in its present form needs many experimental inputs like neutron multiplicity data for all the partitions and initial information of the emitted fragments. It is known that the fission fragment distribution depends on energy and may become symmetric/asymmetric depending on the excitation energy. In our model, this can be achieved by giving different weights to the partitions showing symmetric/asymmetric fragment distribution, which in turn depends on the energy. This information is not present in our model and must come from the measurements or, the detailed theories of fission. The results presented here constitute the first direct evidence of its kind and open up new possibilities for testing the purity of isospin in neutron-rich heavy nuclei and also possibly calculating the fission fragment mass distribution more precisely. More experimental data of this kind will be most useful in further testing and refinement of our model.

\bigskip\noindent
{\large \bf Acknowledgement}
We thank Prof. Don Robson for extensive discussions and valuable comments during the early phases of this work. We also thank Profs. V. K. B. Kota, R. K. Choudhury and P. van Isacker for useful suggestions at various times. Support from Ministry of Human Resource Development (Government of India) to SG and BM in the form of a fellowship is gratefully acknowledged. The authors also acknowledge the financial support from the Department of Science and Technology, Government of India.

\end{document}